\documentclass[twocolumn,eqsecnum,amsmath,showpacs,aps,pre]{revtex4}
\usepackage{psfig}
\begin{document}
%
%

\title{Survival and residence times in disordered chains with bias}

\author{Pedro A. Pury}
\email{pury@famaf.unc.edu.ar}
\affiliation{Facultad de Matem\'atica, Astronom\'\i a y F\'\i sica \\
Universidad Nacional de C\'ordoba \\
Ciudad Universitaria, 5000 C\'ordoba, Argentina}

\author{Manuel O. C\'aceres}
\email{caceres@cab.cnea.gov.ar}
\affiliation{Centro At\'omico Bariloche and Instituto Balseiro \\
Comisi\'on Nacional de Energ\'\i a At\'omica and
Universidad Nacional de Cuyo \\
8400 San Carlos de Bariloche, R\'\i o Negro, Argentina\\
Consejo Nacional de Investigaciones Cient\'\i ficas
y T\'ecnicas, Argentina}

\date{Physical Review E - Received 27 March 2002}

%
%

\begin{abstract}
We present a unified framework for first-passage time and
residence time of random walks in finite one-dimensional disordered
biased systems. The derivation is based on the exact expansion of 
the backward master equation in cumulants.
The dependence on the initial condition, system size, and bias
strength is explicitly studied for models with weak and strong
disorders.
Application to thermally activated processes is also developed.
\end{abstract}

\pacs {05.40.Fb, 05.60.Cd, 02.50.Ga, 66.30.Dn}
\maketitle

\section{Introduction}
\label{sec:Intro}

A large number of physical properties of diffusion and hopping
transport of classical particles (or excitations) in disordered
media have been investigated by means of random walks (RW) in
disordered lattices~\cite{ABSO81,MW87,Reviews}.
Of particular interest are the effects of the finite size and 
boundary conditions on the domain of diffusion. The absorbing 
boundary approach allows us to analyze when a process first 
reaches a given threshold value~\cite{Kam92}.  
This question arises in many situations and is equivalent 
to regarding the relation between the underlying dynamics of randomly
evolving systems and the statisticsof extreme events for such systems.
These statistics are important in a variety of problems in engineering
and applied physics~\cite{LW86}.
Extreme phenomena are experimentally accessible and enable us to
know the parameters of the stochastic dynamics.

One quantity that naturally arises in this context is the time for
which the particle survives before its absorption in the boundary
sinks, i.e., the first-passage time (FPT).  This time depends on 
the realization of the RW, thus being a random variable.
The mean first-passage time (MFPT) is of fundamental importance for
diffusion influenced reactions, as it measures the (reciprocal)
reaction rate constant.
First-passage problems appear in a wide range of
applications~\cite{Kam92} and have a long history~\cite{Sie51}.
Recently, the fact that the MFPT is exactly equal to the inverse of
the associated Kramers escape rate was proved for arbitrary
time-homogeneous stochastic processes~\cite{RSH99}.
Finite size effects also appear when we consider unrestricted
diffusion (no boundary, or boundary condition too far away),
however we ask for the time spent by the diffusing particle
in finite domains. This quantity is the random variable known
as residence time (RT).
Unlike the FPT, which reckons the lifetime of particles that never
abandoned a given domain, the RT involves the case when the particle
can exit from and enter into the domain an unrestricted number of
times.  We must stress that unfortunately the words {\em residence}
and {\em survival} have sometimes been used as synonyms.
We distinguish these terms from the fact that the particle
can return or not (absorption) to the interval of interest.
The mean residence time (MRT) has importance for diffusion
influenced catalytic reactions where reactants are localized
in a finite domain of the catalyzer diffusion region.
Experimental techniques generally called single-molecule
spectroscopy, allow one to follow the evolution in time of
the state of a single molecule that undergoes a conformational
change (isomerization reaction). This fact has recently been
addressed by the MRT study of a single sojourn in each of the
states of the molecule~\cite{SMS}.  The important feature of
this class of reactions is that these can be regarded as being 
one dimensional.
The MRT has found several other applications~\cite{BZ812,AgmBK}.
However, the general problem of RT distribution in random media has
been little discussed in the physics literature~\cite{BZA98}. One of
our goals is to present the FPT and RT statistics in random media in
a unified formalism.

The effect of bias on FPT in disordered systems has received
attention since the first works on the survival fraction of
particles in media with randomly distributed perfect
tramps~\cite{Tramps}.
When the bias is switched on (through external fields), the system
undergoes a substantial change in its dynamics however small the
field is. Several results have been reported on survival probability
(or related quantities) for one-dimensional RW with disorder and
bias.  The models of disorder often regarded are the random traps
and Sinai's model. Trapping in one dimension is a model of strong
disorder that allows exact results with large bias~\cite{ADGSir}.
Sinai's model~\cite{Sinai} is a time discrete RW in one dimension
with asymmetrical transition probabilities that fulfill a certain
condition. In this way, in the Sinai's model the disorder is coupled
with the bias strength.
The FPT problem for Sinai's model has been extensively
studied~\cite{Sinai2}.
In this work we consider a RW on a chain with site disorder 
in the presence of global (site independent) bias. We analyze
weak and strong situations of disorder~\cite{ABSO81,HC90}.
For weak disorder all the inverse moments of the distribution
of RW hopping rates are finite, whereas all these moments diverge
if the disorder is strong.

A successful theory for FPT statistics in disordered media is
the finite effective medium approximation (FEMA)~\cite{HC90}.
FEMA combines an exact expansion of the survival probability
equation in a disordered medium, with the effective medium
approximation~\cite{OL81}. This scheme allows a perturbative
analysis around the effective homogeneous medium in the long
time limit for weak and strong disordered models. FEMA also provides a
self-consistent truncation criterion.  The extension of FEMA to biased
media was presented in Ref.~\cite{Pury94}, where we got perturbative
expansions for small bias and weak disorder.  In Ref.~\cite{CMO+97},
another extension of FEMA was carried out for periodically forced
boundary conditions.
In the present paper, we obtain, in the guidelines of the 
expansions developed in FEMA, the exact equations for MFPT 
and MRT and construct their solutions in the leading relevant 
order for small bias.
MRT for one-dimensional diffusion in a constant field and biased
chains was analyzed in Ref.~\cite{AgmBK}.
However, we could not find explicit expressions for the RT
distribution in disordered media in the literature.
Therefore, another goal of our work is to consider the mixed
effects of disorder and bias in the FPT and RT distributions.

This paper presents the survival and residence times statistics.
In Sec.~\ref{sec:SurRes} we define the survival and residence
probabilities and construct their expressions from the conditional
probability of the random walk on a chain.
The random biased model is described in Sec.~\ref{sec:Random},
whereas in Sec.~\ref{sec:nonDis} the homogeneous (nondisordered)
chain with bias is treated analytically. In Sec.~\ref{sec:Projection}
we introduce the projection operator to average over disorder,
obtaining the main equations. The weak disordered case is analyzed in
Sec.~\ref{sec:weak} and strong disordered cases are considered in
Sec.~\ref{sec:strong}. Finally, in Sec.~\ref{sec:Phys} thermally
activated processes are considered and Sec.~\ref{sec:Fin} provides
the concluding remarks. The mathematical details of the paper are
condensed in two appendixes. In Appendix~\ref{app:Order} the survival
and residence probabilities for homogeneous chains are exactly
calculated, in Appendix~\ref{app:Green} we study the Green's
functions in the presence of bias, and in Appendix~\ref{app:Terw} 
we evaluate the relevant cumulants used in Sec.~\ref{sec:weak}
and~\ref{sec:strong}.

\section{Survival and Residence Times Statistics}
\label{sec:SurRes}

The dynamical behavior of random one-dimensional systems can be
described by the one-step master equation~\cite{ABSO81}
\begin{equation}
\begin{array}{rcl}
\partial_t P(n,t|n_0,t_0)
&=& w_{n-1}^+ \,P(n-1,t|n_0,t_0)  \\
&+ & \;w_{n+1}^- \,P(n+1,t|n_0,t_0)  \\
&- & \,(w_n^+ + w_n^-) \,P(n,t|n_0,t_0) \;\;,
\label{master}
\end{array}
\end{equation}
where $w_n^{+(-)}$ is the transition probability per unit time from
site $n$ to $n+1$ ($n-1$). $P(n,t|n_0,t_0)$ is the conditional
probability of finding the walker at site $n$ at time $t$, given
that it was at site $n_0$ at time $t_0$ ($<t$) and a particular
configuration of $\{w_n^{\pm} \}$.
We assume that $\{ w_n^{\pm} \}$ is a set of positive independent
identically distributed random variables.
In this description, the disorder is modeled by the distribution,
$\rho(w)$, assigned to these random variables.
For a given realization of $\{ w_n^{\pm} \}$ (quenched disorder) we
get a Markovian stochastic dynamics.
We can write Eq.~(\ref{master}) in matrix notation:
$\partial_t {\bf P}(t|t_0) = {\bf H} \;{\bf P}(t|t_0)$, where
\begin{equation}
{\bf H}_{n \,n'} = w_{n'}^+ \;\delta_{n-1 \,n'} +
                   w_{n'}^- \;\delta_{n+1 \,n'} -
                  (w_{n'}^+ + w_{n'}^+) \,\delta_{n \,n'}
\label{H}
\end{equation}
and ${\bf P}(t_0|t_0) = {\bf I}$. Thus, the formal solution of
Eq.~(\ref{master}) is
${\bf P}(t|t_0) = \exp((t - t_0) \,{\bf H})$.
This solution obeys the backward master equation too,
$\partial_t {\bf P}(t|t_0) = {\bf P}(t|t_0) \;{\bf H}$,
for the same initial condition~\cite{Kam92b}.

In this work we consider a RW on a chain and we address the question
about the survival and residence probabilities in the finite interval
$D = [-L,L]$. The first is the probability, $S_{n_0}(t|t_0)$,
of remaining in $D$ (without exiting) at time $t$ if the walker
initially began at site $n_0 \in D$. The second, instead, is the
probability, $R_{n_0}(t|t_0)$, of finding the particle within the
domain $D$ at time $t$, given that it initially began at site $n_0$
(not necessary in $D$).
Therefore, the residence probability is defined by
\begin{equation}
R_{n_0}(t|t_0) = \sum_{n \in D} \,P(n,t|n_0,t_0) \;,
\label{resid}
\end{equation}
where $P(n,t|n_0,t_0)$ is the solution of Eq.~(\ref{master}).
Due to the fact that the master equation links the probabilities
for all sites of the chain, the residence problem involves an
infinite matrix.

To compute the survival probability, we need to eliminate
contributions from trajectories returning to the interval $D$ after
having left it. To do it we must find the solution $P_D(n,t|n_0,t_0)$
of Eq.~(\ref{master}) with absorbing boundaries in the interval's
extremes~\cite{Kam92}.
Thus, $P_D(n,t|n_0,t_0)$ results in the solution of
$\partial_t {\bf P}_D(t|t_0) = {\bf H}_D \;{\bf P}_D(t|t_0)$,
where
\begin{equation}
\left({\bf H}_D \right)_{n \,n'} =
\left\{
\begin{array}{cl}
{\bf H}_{n \,n'} & \mbox{if } n' \in D \\
0                 & \mbox{otherwise} \;.
\end{array}
\right.
\label{H_D}
\end{equation}
Hence, the survival probability results in
\begin{equation}
S_{n_0}(t|t_0) = \sum_{n \in D} \,P_D(n,t|n_0,t_0) \;.
\label{surv}
\end{equation}
This definition and the fact that for the survival problem, we only
need to consider $n_0 \in D$ allow us to work with a finite square
matrix ${\bf H}_D$ of dimension $N \times N$, $N = 2L+1$ being
the number of sites in $D$.

The presented view of the survival problem is adequate for chains
with a fixed number of sites. Nevertheless, we are interested in
general expressions for domains with an arbitrary number $N$ of
sites and we want to work out survival and residence problems
simultaneously.
Due to the time-homogeneous invariance of the problems
we take $t_0=0$ from here and throughout the rest of the work.
We now consider the vector function $F(t)$ whose components are
$F_{n_0}(t) = \sum_{n \in D} \,P(n,t|n_0,0)$.
The evolution equation for this function follows from the
backward master equation and results in
$\partial_t F(t) = {\bf H^{\dag}} \;F(t)$,
where ${\bf H}^{\dag}$ is the transpose matrix of ${\bf H}$.
Using Eq.~(\ref{H}), we can write the last equation in components
\begin{equation}
\begin{array}{rcl}
\partial_t F_{n_0}(t) &=&
w^+_{n_0} \,\left[ F_{n_0+1}(t) - F_{n_0}(t) \right]\\ &+&
w^-_{n_0} \,\left[ F_{n_0-1}(t) - F_{n_0}(t) \right] \;.
\end{array}
\label{EqFt}
\end{equation}
Thus, the residence probability is the solution of
Eq.~(\ref{EqFt}) with the initial condition,
\begin{equation}
R_{n_0}(t=0) =
\left\{
\begin{array}{ll}
1 & \mbox{if } n_0 \in D \\
0 & \mbox{otherwise}
\end{array}
\right.
\end{equation}
and boundary condition at infinity: $R_{n_0}(t) \rightarrow 0$ for
$|n_0| \rightarrow \infty$ for all finite $t$.
On the other hand, the survival probability is the solution of the
generic adjoint equation~(\ref{EqFt}) for the infinite chain with
the initial condition $S_{n_0}(t=0)=1$ for all $n_0 \in D$.
Here, the {\em artificial} boundary condition $S_{n_0}(t) = 0$ for
all $t$ if $n_0 = -(L+1)$ or $n_0 = L+1$ must be used to prevent the
back flow of the probability into the interval~\cite{note:onestep}.

The survival probability decreases monotonically in time from unity
to zero.  Let us now introduce the first-passage time distribution
(FPTD) $f_{n_0}(t)$, i.e., the probability density of exit $D$ at
a time between $t$ and $t+dt$; then
$f_{n_0}(t) = - \partial_t S_{n_0}(t)$~\cite{Kam92}.
The MFPT is the first moment $T_{n_0}$ (if it exists) of FPTD,
\begin{equation}
T_{n_0} = \int_0^{\infty} t \,f_{n_0}(t) \,dt \;.
\label{T_f}
\end{equation}
If $t \,S_{n_0}(t) \rightarrow 0$ for $t \rightarrow \infty$,
then
\begin{equation}
T_{n_0} = \int_0^{\infty} S_{n_0}(t) \,dt \;.
\label{T_S}
\end{equation}
The residence probability does not necessarily decrease to zero at
infinitely long times. Moreover, it need not even be monotonic in
time. Thus, the residence time density is generally not equal to the
negative time derivative of the residence probability~\cite{BZA98}.
Nevertheless, we can define the MRT, $\tau_{n_0}$, in a manner
analogous to Eq.~(\ref{T_S}), namely,
\begin{equation}
\tau_{n_0} = \int_0^{\infty} R_{n_0}(t) \,dt \;.
\label{tau_R}
\end{equation}
Thus, from Eqs.~(\ref{T_S}) and~(\ref{tau_R}),
we obtain MFPT and MRT from the asymptotic limit
of the Laplace transformed (denoted by hats)
survival and residence probabilities~\cite{note:MFPT},
\begin{subequations}
\begin{eqnarray}
T_{n_0} &=& \lim_{s \rightarrow 0} \hat{S}_{n_0}(s) \;,
\label{T_Ss}
\\
\tau_{n_0} &=& \lim_{s \rightarrow 0} \hat{R}_{n_0}(s) \;.
\label{tau_Rs}
\end{eqnarray}
\label{T_Fs}
\end{subequations}
For MFPT, this limit exists if
$\hat{S}_{n_0}(s) \sim T_{n_0} + c \,s^{\chi}$,
where $c$ and $T_{n_0}$ are assumed constants and $\chi > 0$.
In this manner, from $\hat{f}_{n_0}(s) = 1 - s \, \hat{S}_{n_0}(s)$,
the normalization condition of the FPTD is also guaranteed:
$\int_0^{\infty} f_{n_0}(t) \,dt = \lim_{s \rightarrow 0}
\hat{f}_{n_0}(s) = 1$.

\section{Random Biased Model}
\label{sec:Random}

In this work, we are interested in the interplay between the bias and
the disorder in the transition probabilities. For this goal we take
\begin{equation}
w_n^+ = a + \xi_n \;\;\; , \;\;\; w_n^- = b + \xi_n \;,
\label{model}
\end{equation}
where $a$ and $b$ are positive constants and $\left\{ \xi_n \right\}$
are taken to be independent but identically distributed random
variables with $\langle \xi_n \rangle = 0$.
This form of jump transitions involves an ordered biased background
with a superimposed random medium. The strength of the bias is given
by the ratio between $a$ and $b$ and the disorder is characterized
by the distribution of variables $\left\{ \xi_n \right\}$.
Without lost of generality we assume $a \geq b$ and in consequence
we must to impose the restriction $\xi_n \geq -b$. This restriction
guarantees the positivity of jump probabilities $\{ w_n^{\pm} \}$.
We introduce the parameter $\epsilon$ for bias strength by
$b/a = 1 - \epsilon$ and we take $0 \leq \epsilon \leq 1$, so that
the bias field points to the right.
This election of parameters allows us to focus our attention in
the small bias limit and to study the transition to the symmetric
diffusive behavior~\cite{note:multipl}.
The Laplace transform of the evolution equation for our model results
from Eq.~(\ref{EqFt}),
\begin{equation}
s \hat{F}_n(s) - F_n(t=0) =
\left[ {\cal K}^b + \xi_n \, {\cal K}^0 \right] \,\hat{F}_n(s) \;,
\label{EqFs}
\end{equation}
where we have introduced the operators
\begin{equation}
\begin{array}{rcl}
{\cal K}^0 &\equiv& {\cal E}^+ + {\cal E}^- - 2 \; {\cal I} \;, \\
{\cal K}^b &\equiv& a \; ({\cal E}^+ - {\cal I})
                  + b \; ({\cal E}^- - {\cal I}) \;.
\end{array}
\label{operators}
\end{equation}
${\cal E}^{\pm}$ are shifting operators
(${\cal E}^{\pm} g_n \equiv g_{n \pm 1}$) and
${\cal I}$ is the identity operator.
Equation (\ref{EqFs}) must be solved with the boundary conditions
corresponding to each problem,
\begin{subequations}
\begin{eqnarray}
\hat{S}_{-(L+1)}(s) = \hat{S}_{(L+1)}(s) = 0 \,,
\label{Sboundary}
\\
\hat{R}_{n}(s) \rightarrow 0 \mbox{ for } |n| \rightarrow \infty \,,
\label{Rboundary}
\end{eqnarray}
\label{Fboundary}
\end{subequations}
$\forall \;s > 0$. The initial condition is given by
\begin{equation}
F_{n}(t=0) =
\left\{
\begin{array}{ll}
1 & \mbox{if } n \in D \\
0 & \mbox{otherwise} \;.
\end{array}
\right.
\label{icondition}
\end{equation}
Remember that for FPT we only need to consider $n \in D$.

The classes of disorder analyzed are generalizations of
standard cases in the literature~\cite{ABSO81,HC90}.
Our expressions are constructed introducing the parameter $\epsilon$
in such a way that we guarantee the positivity of transition rates and
reproduce the known expressions in the limit
$\epsilon \rightarrow 0$.
We have considered the following three classes of disorder for the
transfer rate $w = w_n^+$.
\renewcommand{\theenumi}{\alph{enumi}}
\begin{enumerate}
\item The mean values of the inverse moments of jump transition $w$,
$\beta_k \equiv \langle \left( 1/w \right)^k \rangle$,
are finite quantities for all $k \geq 1$, $\epsilon > 0$ and
remain finite in the limit $\epsilon \rightarrow 0$.
\label{weak}
\item The probability distribution $\rho(w)$ is
\begin{equation}
\rho(w) =
\left\{
\begin{array}{cl}
B & \mbox{if } w \in (a \,\epsilon, \Omega_B) \\
0 & \mbox{otherwise} \;,
\end{array}
\right.
\label{model_b}
\end{equation}
where the values of $B$ and $\Omega_B$ are fixed
by the normalization condition
($\int_0^{\infty} \rho(w) \,dw = 1$)
and the fact that $\langle w \rangle = a$,
\begin{subequations}
\begin{eqnarray}
B &=& \left( 2 \,(1-\epsilon) \,a \right)^{-1} \;,
\\
\Omega_B &=& (2-\epsilon) \,a \;.
\end{eqnarray}
\label{normaB}
\end{subequations}
\label{windows}
\item The probability distribution $\rho(w)$ is
\begin{equation}
\rho(w) =
\left\{
\begin{array}{cl}
C \,w^{-\alpha}
& \mbox{if } w \in (a \,\epsilon, \Omega_C) \\
0 & \mbox{otherwise} \;,
\end{array}
\right.
\label{model_c}
\end{equation}
where $0 < \alpha < 1$ and the values of $C$ and $\Omega_C$ are also
fixed by the normalization condition and $\langle w \rangle = a$.
For small $\epsilon$ it gives
\begin{subequations}
\begin{equation}
C \approx \frac{(1-\alpha)^{2-\alpha}}
{\left[ (2-\alpha) \,a \right]^{1-\alpha}}
\left[ 1 + (2-\alpha)^{\alpha}
\,\left( (1-\alpha) \,\epsilon \right)^{1-\alpha} \right] ,
\end{equation}
\begin{equation}
\Omega_C \approx \frac{2-\alpha}{1-\alpha} \,a
\,\left[ 1 - \left( \frac{1-\alpha}{2-\alpha} \right)^{1-\alpha}
\,\epsilon^{1-\alpha} \right] \;.
\end{equation}
\label{normaC}
\end{subequations}
\label{alpha}
\end{enumerate}
The expressions given in Eqs.~(\ref{model_b}) and (\ref{normaB})
can be obtained from the corresponding expressions given
in Eqs.~(\ref{model_c}) and (\ref{normaC}) in the limit
$\alpha \rightarrow 0$.
Class (\ref{weak}) corresponds to the situation of
weak disorder. There, the mean-square displacement of
the RW behaves like $t$ for long times.
Classes (\ref{windows}) and (\ref{alpha}) become strong
disordered cases for $\epsilon \rightarrow 0$, and correspond to
situations of anomalous diffusion.
For class (\ref{windows}),
$\beta_1 \propto \ln (1 / \epsilon)$
and $\beta_k \propto \epsilon^{1-k}$ if $k>1$.
For class (\ref{alpha}),
$\beta_k \propto \epsilon^{1-k-\alpha}$.
In the strong disorder limit, $\langle n(t)^2 \rangle$ behaves
for long times as $t/\ln t$ and $t^{2(1-\alpha)/(2-\alpha)}$,
for cases (\ref{windows}) and (\ref{alpha}), respectively.
Though, the MFPT in the presence of strong disorder is a divergent
quantity~\cite{HC90}. We will show in Sec.~\ref{sec:strong}
that our model allows us to study the transition to strong
disorder in the limit $\epsilon$ going to zero, i.e.,
the zero bias limit.

\section{Nondisordered Biased Chain}
\label{sec:nonDis}

The {\em nondisordered} case is obtained from the trivial
distribution $\rho(w) = \delta(w - a)$.
Therefore, basic results about the survival and residence
probabilities in nondisordered chains can be easily obtained
from the equation:
$s \hat{F}_n(s) - F_n(0) = {\cal K}^b \hat{F}_n(s)$,
with the boundary conditions given by Eq.~(\ref{Fboundary}).
In particular, exact expressions for MFPT and MRT for a homogeneous
biased chain are given by (see Appendix~\ref{app:Order} for detailed
calculations)
\begin{equation}
T_n = \frac{L+1-n}{a(1-\gamma)} - \frac{2 \,(L+1)}{a(1-\gamma)}
\,\frac{\gamma^n - \gamma^{L+1}}{\gamma^{-(L+1)} - \gamma^{L+1}}
\,,
\label{TordL}
\end{equation}
with $-L \leq n \leq L$, and
\begin{equation}
\tau_n = \frac{1}{a} \left\{
\begin{array}{ll}
\displaystyle
\frac{2\,L+1}{1-\gamma}
& , \, n < -L \\
\displaystyle
\frac{L-n}{1-\gamma} + \frac{1-\gamma^{n+L+1}}{(1-\gamma)^2}
& , \, -L \leq n \leq L \\
\displaystyle
\frac{1-\gamma^{2L+1}}{(1-\gamma)^2} \;\gamma^{n-L}
& , \, n >  L \,,
\end{array}
\right.
\label{TauordL}
\end{equation}
%
\begin{figure}
\psfig{file=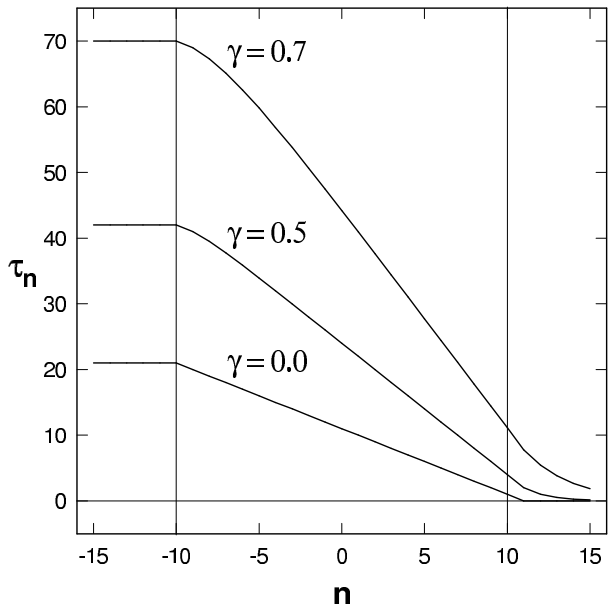,width=80mm}
\caption{
MRT for a nondisordered chain [as given by Eq.~(\ref{TauordL})]
plotted against the discrete initial condition~$n$, with~$a=1$
and~$L=10$. The solid lines are only to guide the eye.
}
\label{FigOrd}
\end{figure}
%
where $\gamma = b/a < 1$.
Figure~\ref{FigOrd} shows the behavior
of~$\tau_n$ for some values of~$\gamma$.
We would like to stress that for $n<-L$, MRT is a constant
proportional to the width of the interval ($2\,L+1$),
whereas for $n \rightarrow \infty$, MRT vanishes.
Given that the bias points to the right, for any initial condition
at the left of the interval of interest, MRT is equal to the
transit time across the interval. For a given initial condition,
MRT is greater whereas the bias is smaller.
For one way motion ($\gamma=0$), from Eq.~(\ref{TauordL})
we obtain
\begin{equation}
\tau_n = \frac{1}{a} \left\{
\begin{array}{ll}
2\,L+1
& , \; n < -L \\
L + 1 - n
& , \; n \in D \\
0
& , \; n >  L \,,
\end{array}
\right.
\label{TauDirect}
\end{equation}
whereas in the small bias limit, i.e., $\gamma=1-\epsilon$,
$\epsilon << 1$, taking $a$ constant results in
\begin{equation}
\tau_n = \frac{2\,L+1}{a\,\epsilon} \left\{
\begin{array}{ll}
1
& \,, \, n < -L \\
\displaystyle
1 - \frac{(L+n)\,(L+n+1)}{2\,(2\,L+1)} \,\epsilon
& \,, \, n \in D \\
1 - n \,\epsilon
& \,, \, n >  L \,.
\end{array}
\right.
\label{TauDiff}
\end{equation}

It is worthwhile to emphasize that MFPT exhibits a crossover from
the drift (strong bias) regime [$T_n = (L+1-n)/a)$] to the diffusive
(small bias) regime ($T_n \propto ((L+1)^2-n^2)/2a$). The diffusive
behavior is also present for MRT for~$n \in D$, but it is not in the
leading term.
Moreover, MFPT remains finite in the limit~$\epsilon \rightarrow 0$
for finite domains, whereas MRT diverges as we can see from
Eq.~(\ref{TauDiff}).
Thus, the MRT is not a defined quantity for unbiased
chains~\cite{note:MRT}.
Expressions for the MFPT in nondisordered biased chains
following from Eq.~(\ref{TordL}), and the study of the drift
and diffusive regimes were reported in Ref.~\cite{Pury94}.

\section{Projection Operator Average}
\label{sec:Projection}

The basic equation for the evolution of probabilities has been
written in Eq.~(\ref{EqFs}). The operators in this equation
explicitly show the splitting of the transition probability
in an average biased part
($\langle w_n^+ \rangle = a$, $\langle w_n^- \rangle = b$)
and a random nonbiased part ($\xi_n$).
Defining the operator $\Delta = \xi_n \,{\cal K}^0$, let us rewrite
Eq.~(\ref{EqFs}) as
\begin{equation}
s \,\hat{F}(s) - F(0) =
\left[ {\cal K}^b + \Delta \right] \,\hat{F}(s) \;.
\label{EqFdelta}
\end{equation}
Our goal in this section is to obtain exact equations for the
averaged survival and residence probabilities.
This average can be formally carried out introducing a projection
operator $\cal P$ (${\cal P}^2 = {\cal P}$) that averages over the
joint probability density of variables $\{\xi_n\}$:
$\langle\hat{F}\rangle \equiv {\cal P} \,\hat{F}$,
$\hat{F} = \langle\hat{F}\rangle + (1 - {\cal P}) \,\hat{F}$.
Applying the operator $\cal P$ to Eq.~(\ref{EqFdelta}) we obtain
\begin{equation}
s \,\langle\hat{F}\rangle - F(0) =
{\cal K}^b \langle\hat{F}\rangle +
{\cal P} \Delta \langle\hat{F}\rangle +
{\cal P} \Delta (1 - {\cal P}) \hat{F} \;.
\label{PF}
\end{equation}
Also, applying the operator $(1 - {\cal P})$ to Eq.~(\ref{EqFdelta})
we arrive to
\begin{equation}
\begin{array}{rcl}
s (1 - {\cal P}) \hat{F} &=& {\cal K}^b (1 - {\cal P}) \hat{F} +
(1 - {\cal P}) \Delta \langle\hat{F}\rangle \\
&+& (1 - {\cal P}) \Delta (1 - {\cal P}) \hat{F} \;.
\end{array}
\label{(1-P)F}
\end{equation}
A formal solution of Eq.~(\ref{(1-P)F}) can be obtained using the
Green's function for the nondisordered chain,
\begin{equation}
\hat{G}(s) = \left( s - {\cal K}^b \right)^{-1} \;.
\label{G}
\end{equation}
Applying $\hat{G}$ to Eq.~(\ref{(1-P)F}) and using the definition
given in Eq.~(\ref{G}), results in
\begin{equation}
(1 - {\cal P}) \hat{F} = \hat{G}
\left[ (1 - {\cal P}) \Delta \langle\hat{F}\rangle +
(1 -{\cal P}) \Delta (1 -{\cal P}) \hat{F} \right] \;.
\label{(1-P)F_for}
\end{equation}
Equation~(\ref{(1-P)F_for}) can be iteratively solved for
$(1 - {\cal P}) \hat{F}$,
\begin{equation}
(1 - {\cal P}) \hat{F} = \sum_{k=1}^{\infty}
\left[ \hat{G} (1 - {\cal P}) \Delta \right]^k
\langle\hat{F}\rangle
\;.
\label{(1-P)F_ite}
\end{equation}
Putting this formal solution in Eq.~(\ref{PF}) we find
a closed exact equation for the average probability
$\langle\hat{F}\rangle$,
\begin{equation}
s \,\langle\hat{F}\rangle - F(0) =
{\cal K}^b \langle\hat{F}\rangle +
\left\langle \sum_{k=0}^{\infty}
\left[ \Delta \hat{G} (1 - {\cal P}) \right]^k \Delta
\right\rangle \langle\hat{F}\rangle \,.
\label{avF0}
\end{equation}
The operator $\hat{G}$ is solution of the equation,
\begin{equation}
\left( s - {\cal K}^b \right) \hat{G} = \openone \,,
\label{EqG}
\end{equation}
with the boundary conditions
\begin{subequations}
\begin{equation}
\hat{G}^S_{-(L+1) \,m}(s) = \hat{G}^S_{(L+1) \,m}(s) = 0
\;\forall \,s \mbox{ and } m \in D \;,
\label{GSboundary}
\end{equation}
\begin{equation}
\hat{G}^R_{n m}(s) \rightarrow 0 \;
\mbox{ for } |n| \rightarrow \infty
\mbox{ and } m \mbox{ finite} \;,
\label{GRboundary}
\end{equation}
\label{Gboundary}
\end{subequations}
where the superscript~$S$ ($R$) corresponds to the survival
(residence) problem.
Exact expressions of the Green's functions are calculated
in the Appendix~\ref{app:Green}.

We will find it useful to write Eq.~(\ref{avF0}) in components.
For this task, we use the explicit form of $\Delta$, the Terwiel's
cumulants~\cite{Terwiel} of the random variables $\xi_k$:
$\langle \xi_n \xi_{m_1} \ldots \xi_{m_p} \rangle_T =
{\cal P} \,\xi_{n} \,(1 - {\cal P}) \,\xi_{m_1}
\dots (1 - {\cal P}) \,\xi_{m_p}$,
and we define the propagator:
$J_{n m}(s) = {\cal K}^0 \,\hat{G}_{n m}(s)$, where the operator
${\cal K}^0$ acts on the first index of $\hat{G}_{n m}(s)$.
Explicit expressions for $J_{n m}(s)$ are given by Eqs.~(\ref{JS})
and~(\ref{JR}) in Appendix~\ref{app:Green}.
In this manner we can write
\begin{eqnarray}
s \,\langle\hat{F}_n\rangle - F_n(0) &=&
{\cal K}^b \langle\hat{F}_n\rangle
\,+\, \sum_{p=0}^{\infty} \, \sum_{m_1, \ldots, m_p}
\nonumber \\
&&\times
\langle \xi_n \xi_{m_1} \ldots \xi_{m_p} \rangle_T
\, J_{n m_1}(s)
\nonumber \\
&&\times
J_{m_1 m_2}(s) \ldots J_{m_{p-1} m_p}(s) \;
{\cal K}^0 \langle\hat{F}_{m_p}\rangle \,.
\nonumber \\
\label{avF1}
\end{eqnarray}
We must understand that $m_0 = n$, and for the survival
problem we have the additional restriction $m_1, \ldots, m_p \in D$.
The exact effective backward equation given by Eq.~(\ref{avF1}) can
be rewritten as
\begin{eqnarray}
s \,\langle\hat{F}_n\rangle - F_n(0) &=&
{\cal K}^b \langle\hat{F}_n\rangle
\,+\, \sum_{p=0}^{\infty} \,
\sum_{  {{m_1 \neq n \atop m_2 \neq m_1} \atop \vdots}
\atop m_p \neq m_{p-1}  }
\nonumber \\
&&\times
\langle \psi_n \psi_{m_1} \ldots \psi_{m_p} \rangle_T \;
\, J_{n m_1}(s)
\nonumber \\
&&\times
J_{m_1 m_2}(s) \ldots J_{m_{p-1} m_p}(s) \;
{\cal K}^0 \,\langle\hat{F}_{m_p}\rangle \,,
\nonumber \\
\label{avF2}
\end{eqnarray}
where we have used the definition of Terwiel's cumulants.
Here, we have summed up all the terms containing the diagonal parts
of $J_{n m}(s)$ through the introduction of the random operator
$\psi_k(s)$ defined by
\begin{equation}
\psi_k(s) = \sum_{i_k=0}^{\infty}
\,\left[ J_{k k}(s) \,\xi_k (1 - {\cal P})  \right]^{i_k} \,\xi_k \,.
\label{psi1}
\end{equation}
This operator acts on any disorder-dependent quantity at its right.
The geometrical sum in Eq.~(\ref{psi1}) can be evaluated, resulting
in
\begin{equation}
\psi_k(s) = M_k(s) -
\frac{M_k(s) \,J_{k k}(s)}{1 + \langle M_k(s) \rangle \,J_{k k}(s)}
\,{\cal P} M_k(s) \,,
\label{psi2}
\end{equation}
where
\begin{equation}
M_k(s) = \frac{\xi_k}{1 - \xi_k \,J_{k k}(s)} \,.
\label{M}
\end{equation}

In the following, we take the limit $s \rightarrow 0$ in order to
obtain from Eq.~(\ref{avF2}) the corresponding equations for the
MFPT and MRT in disordered media, which are defined from
Eq.~(\ref{T_Fs}) by
\begin{subequations}
\begin{eqnarray}
\left< T_{n} \right> &=&
\lim_{s \rightarrow 0} \left< \hat{S}_{n}(s) \right> \;,
\label{<T>}
\\
\left< \tau_{n} \right> &=&
\lim_{s \rightarrow 0} \left< \hat{R}_{n}(s) \right> \;.
\label{<tau>}
\end{eqnarray}
\label{<Ttau>}
\end{subequations}
In this limit the propagator $J_{n m}(s)$ can be written as
$\Lambda_{n m} + {\cal O}(s)$,
and the exact expression for $\Lambda_{n m}$ in the FPT (RT)
problem is given by Eq.~(\ref{LambdaS}) [Eq.~(\ref{LambdaR})]
in Appendix~\ref{app:Green}. Therefore, the resulting equation
for the averaged MFTP ($\langle T_n \rangle$) is
\begin{eqnarray}
- S_n(t=0) &=&
{\cal K}^b \langle T_n \rangle
\,+ \sum_{p=0}^{\infty} \,
\sum_{  {{m_1 \neq n \atop m_2 \neq m_1} \atop \vdots}
\atop m_p \neq m_{p-1}  }
\langle \psi_n \psi_{m_1} \ldots \psi_{m_p} \rangle_T^S
\nonumber \\
&& \times
\Lambda^S_{n m_1} \Lambda^S_{m_1 m_2} \ldots \Lambda^S_{m_{p-1} m_p}
\;{\cal K}^0 \,\langle T_{m_p} \rangle \,.
\nonumber \\
\label{avTeq}
\end{eqnarray}
Its solution must satisfy the boundary conditions:
$\langle T_{-(L+1)} \rangle = \langle T_{L+1} \rangle = 0$.
A similar equation for the averaged MRT ($\langle \tau_n \rangle$)
is obtained from Eq.~(\ref{avTeq}) replacing the $S$ quantities by
the corresponding $R$ quantities and imposing the boundary conditions:
$\langle \tau_n \rangle = $ const for $n < -L$ and
$\langle \tau_n \rangle \rightarrow 0$ for $n \rightarrow +\infty$
(given that the bias field points to right).

We, additionally, consider the case of small bias.
Thus, the expressions for $\Lambda_{n m}$ can be further
expanded, and taking $a$ constant results in
\begin{subequations}
\begin{equation}
\Lambda^S_{n m} \approx
\left\{
\begin{array}{ll}
\displaystyle
\frac{1}{2 \,(L+1) \,a} \,[m - (L+1)] \, \epsilon
&, n < m\\
\displaystyle
-\frac{1}{a} +
\frac{1}{2 \,(L+1) \,a} \,[m - (L+1)] \, \epsilon
&, n = m\\
\displaystyle
\frac{1}{2 \,(L+1) \,a} \,[m + (L+1)] \, \epsilon
&, n > m \;,
\end{array}
\right.
\end{equation}
\begin{equation}
\Lambda^R_{n m} \approx
\frac{1}{a}
\left\{
\begin{array}{rl}
0
&, n < m\\
-1
&, n = m\\
\epsilon
&, n > m \;.
\end{array}
\right.
\end{equation}
\label{Lambda}
\end{subequations}
Again, the superscript~$S$ ($R$) corresponds to the survival
(residence) problem.
Hence, $\Lambda_{n m} \propto \epsilon$ for $n \neq m$ and,
from Eq.~(\ref{LambdaR}), the diagonal components of the propagator
in the residence problem result independent of $\epsilon$.

\section{Weak Disorder}
\label{sec:weak}

For disorder of class~(\ref{weak}), the quantities $\beta_k$ are
finite and we obtain that
$\langle \psi_n \psi_{m_1} \ldots \psi_{m_p} \rangle_T^S
\propto \epsilon^0$   and
$\langle \psi_n \psi_{m_1} \ldots \psi_{m_p} \rangle_T^R$
is independent of $\epsilon$.
Thus, Eq.~(\ref{avTeq}) is a perturbative expansion in the sense
that $\langle T_n \rangle$ can be calculated up to order $\epsilon^q$
truncating the $p$ series in the $q$ term ($p = 0, \dots, q$).
The corresponding equation for $\langle \tau_n \rangle$ is strictly
a perturbative expansion given that the contribution to the order
$\epsilon^q$ comes entirely from the term $p=q$.

In Eq.~(\ref{avTeq}), the cumulant for $p=1$ consists of
two independent random operators, so it vanishes
(see Appendix~\ref{app:Terw}).
Therefore, it turns out that only the term with $p=0$
contributes to order $\epsilon$.
From Eq.~(\ref{T13weak}) results
\begin{equation}
\langle \psi_n \rangle_T^S =
\beta_1^{-1} - a - a \,{\cal F} \,\frac{L+1-n}{2 \,(L+1)} \,\epsilon
\;,
\label{T1Sweak}
\end{equation}
where we have introduced the fluctuation of the quenched disorder:
${\cal F} = (\beta_2 - \beta_1) / \beta_1^2$.
Up to order $\epsilon$, taking $a$ constant and using
${\cal K}^b \approx
a \left[ {\cal K}^0 -  ({\cal E}^- - {\cal I}) \,\epsilon \right]$,
the explicit form of Eq.~(\ref{avTeq}) is
\begin{eqnarray}
- 1 &=& \beta_1^{-1} \, {\cal K}^0 \,\langle T_n \rangle
\nonumber \\
&&
- a \left[ {\cal F} \,\frac{L+1-n}{2 \,(L+1)} \, {\cal K}^0
+ ({\cal E}^- - {\cal I}) \right] \epsilon \,\langle T_n \rangle
+ {\cal O}(\epsilon^2) \,.
\nonumber \\
\label{MFPTeq}
\end{eqnarray}
For $\epsilon = 0$, Eq.~(\ref{MFPTeq}) immediately gives the
well-known MFPT for the unbiased case:
$\langle T_n(\epsilon=0) \rangle =
\left[ (L+1)^2 - n^2 \right] / \left( 2 \,\beta_1^{-1} \right)$,
where we can see that the effect of weak disorder is to replace
the constant $a$ by the effective coefficient $\beta_1^{-1}$.
To construct the consistent solution up to order $\epsilon$
of Eq.~(\ref{MFPTeq}), we propose the expression
\begin{equation}
\langle T_n \rangle = \frac{(L+1)^2 - n^2}{2 \,\beta_1^{-1}}
\left[ 1 + (A \,n + B) \,\epsilon \right] \,,
\label{SolMFPT}
\end{equation}
which immediately satisfies the boundary conditions:
$\langle T_{-(L+1)} \rangle = \langle T_{L+1} \rangle = 0$.
To fit the constants $A$ and $B$ we substitute this
expression in Eq.~(\ref{MFPTeq}) and retain only the
terms up to order $\epsilon$.
If the factor of $\epsilon$ in the expression~(\ref{SolMFPT}) were
a polynomial of degree greater than $1$, we can easily see that
the coefficients of the terms of degree greater than $1$ vanish.
Thus, we obtain for $n \in D$
\begin{eqnarray}
\langle T_n \rangle &\approx& \frac{(L+1)^2 - n^2}{2 \,\beta_1^{-1}}
\nonumber \\
&& \times
\left[ 1 + \left( \frac{3-2n}{6} +
\frac{L+1-n/3}{2 \,(L+1)} \,{\cal F} \right)
a \beta_1 \,\epsilon \right] .
\nonumber \\
\label{MFPT}
\end{eqnarray}
From this expression we can analyze the interplay between the
bias and the fluctuation of the quenched disorder ${\cal F}$.
In the presence of bias, the escape time from the finite interval
increases for all initial conditions $n$, with respect to the
unbiased case, if the fluctuation is large enough.
This is an important result related to the control of the
trapping process. This fact was reported in Ref.~\cite{Pury94}
where the solution of Eq.~(\ref{MFPTeq}) was constructed from FEMA.
Now, we can evaluate the difference between the exact solution
(order $\epsilon$) given by Eq.~(\ref{MFPT}) and that approximation,
obtaining that the correction to FEMA is
$\frac{(L+1)^2 - n^2}{6 \,(L+1)} \,n \,{\cal F} \,a \,\epsilon$.

In the residence problem, from Eqs.~(\ref{theta})
and~(\ref{T13weak}) results
$\langle \psi_n \rangle_T^R = \beta_1^{-1} - a$.
Thus, the corresponding equation for the averaged MRT is
\begin{equation}
- R_n(0) = \beta_1^{-1} \, {\cal K}^0 \,\langle \tau_n \rangle
- a \,\epsilon \,({\cal E}^- - {\cal I}) \,\langle \tau_n \rangle
+ {\cal O}(\epsilon^2) \,,
\label{MRTeq}
\end{equation}
where $R_n(0)$ is given by Eq.~(\ref{icondition}).
Now we have to deal with a backward master equation with constant
coefficients.
In this case, {\em the fluctuations of disorder are not present}.
Formally, this equation is equal to the one corresponding to a
nondisordered chain in the small limit bias.
Therefore, from Eq.~(\ref{TauDiff}), with the substitutions
\begin{eqnarray}
a &\rightarrow& \beta_1^{-1} \,,
\nonumber \\
b &\rightarrow& \beta_1^{-1} (1 - a \,\beta_1 \,\epsilon) \,,
\nonumber \\
\epsilon &\rightarrow& a \,\beta_1 \,\epsilon \,,
\label{map1}
\end{eqnarray}
we obtain, up to order $\epsilon$, the solution
\begin{equation}
\tau_n \approx \frac{2\,L+1}{a\,\epsilon} \left\{
\begin{array}{ll}
1
& , n < -L \\
\displaystyle
1 - \frac{(L+n)\,(L+n+1)}{2\,(2\,L+1)} a \beta_1 \epsilon
& , n \in D \\
1 - n \,a \,\beta_1 \,\epsilon
& , n >  L \;.
\end{array}
\right.
\label{MRT}
\end{equation}
%

\section{Strong Disorder}
\label{sec:strong}

For the classes of strong disorder, $\beta_k$ are divergent
quantities in the limit $\epsilon \rightarrow 0$,
\begin{subequations}
\begin{equation}
\begin{array}{ll}
\beta_1 \approx |\ln \epsilon| / (2a) & \\
\beta_k \approx \displaystyle\frac{(a \epsilon)^{1-k}}{2a(k-1)}
& (k > 1)
\end{array}
,\;\mbox{for class (\ref{windows})},
\end{equation}
\begin{equation}
\beta_k \approx \frac{C}{k-1+\alpha}(a \epsilon)^{1-k-\alpha}
,\;\mbox{for class (\ref{alpha})},
\end{equation}
\label{betas}
\end{subequations}
where $0 < \alpha < 1$ and $C$ is given in Eq.~(\ref{normaC}).
Also in this limit, Terwiel's cumulants diverge except the
first one, $\langle \psi_n \rangle_T$, which vanishes.
However, all the terms of the $p$ series in the
Eq.~(\ref{avTeq}) vanish and the one corresponding to $p=0$
is the leading term. We reckon the relevant cumulants
in Appendix~\ref{app:Terw}.

For a disorder of class (\ref{windows}) we obtain
${\cal F} \approx \frac{2}{\epsilon \,\ln^2 \epsilon}$,
and from Eq.(\ref{T1Sweak}) results
$\langle \psi_n \rangle_T^S \approx \beta_1^{-1}$.
Therefore, the leading terms in the equation for the
averaged MFPT result in
\begin{equation}
- 1 = \beta_1^{-1} \,{\cal K}^0 \,\langle T_n \rangle
+ {\cal O} \left( \frac{\epsilon}{|\ln \epsilon|} \right) \,.
\label{eq-windows}
\end{equation}
This is a backward master equation for the unbiased chain
with constant coefficients whose solution is given by
$\langle T_n \rangle =
\left[ (L+1)^2 - n^2 \right] / \left( 2 \,\beta_1^{-1} \right)$.
Hence, the averaged MFPT diverges as
$\beta_1 \propto |\ln \epsilon|$.

On the other hand, for a disorder of class (\ref{alpha}) we obtain
${\cal F} \approx \frac{1}{C} \frac{\alpha^2}{1 + \alpha}
(a \,\epsilon)^{\alpha-1}$,
and from Eq.~(\ref{T1Sweak}) results
$\langle \psi_n \rangle_T^S \approx \beta_1^{-1}
\left( 1 - \frac{\alpha}{1 + \alpha}
\,\frac{L+1-n}{2 \,(L+1)} \right)$.
Therefore, the MFPT's equation is
\begin{equation}
- 1 = \left( 1 - \frac{\alpha}{1 + \alpha}
\,\frac{L+1-n}{2 \,(L+1)} \right)
\beta_1^{-1} \,{\cal K}^0 \,\langle T_n \rangle
+ {\cal O}(\epsilon) \,.
\label{eq-alpha}
\end{equation}
In this case, we get an unbiased backward master equation with
linear coefficients. General solutions of this equation will be
given elsewhere. Nevertheless, we hold our attention in the divergent
behavior of the averaged MFPT for $\epsilon \rightarrow 0$.
From Eq.~(\ref{eq-alpha}) we can see that
$\langle T_n \rangle \propto \beta_1 \propto \epsilon^{-\alpha}$.

FEMA~\cite{Pury94} consists in introducing an effective
nonhomogeneous medium $\Gamma_n$ and truncating Eq.~(\ref{avTeq})
to the first term:
$-1 = {\cal K}^b \langle T_n \rangle$.
Here, the operator ${\cal K}^b$ is constructed with the constants
$a + \Gamma_n$ and $b + \Gamma_n$.
The effective rate, $\Gamma_n$,
is fixed imposing the condition
$\langle \psi_n(\Gamma_n) \rangle_T^S = 0$.
The resulting solutions give the exact laws:
$\langle T_n \rangle \propto |\ln \epsilon|$,
for a disorder of class (\ref{windows}) and
$\langle T_n \rangle \propto \epsilon^{-\alpha}$,
for a disorder of class (\ref{alpha}), however in the last case
the predicted $n$-dependent coefficient is not exact.
Strikingly, the predicted divergence laws for the averaged 
MFPT in biased chains agree with the corresponding laws for 
the survival probability, obtained for the unbiased case in 
the limit $s \rightarrow 0$~\cite{HC90}.
The behavior of the survival probability obtained from FEMA is
$\langle \hat{S}_n(s) \rangle \propto |\ln s|$,
for a disorder of class (\ref{windows}), and
$\langle \hat{S}_n(s) \rangle \propto s^{-\alpha}$,
for a disorder of class (\ref{alpha}).
Accordingly, both problems have the same exponents
despite their different nature.

In the residence problem, for both classes of strong disorder
we get that $\langle \psi_n \rangle_T^R \approx \beta_1^{-1}$,
and then the equation for the averaged MRT results
$- R_n(0) = \beta_1^{-1} \,{\cal K}^0 \,\langle \tau_n \rangle$,
i.e., an unbiased equation. Therefore, the averaged MRT
is not a defined quantity for strong disorder.

\section{Thermally Activated Processes}
\label{sec:Phys}

Before concluding this work, we want to analyze the physical
realization corresponding to thermally activated processes in
weakly disordered chains.
For this goal we suppose a particle that moves in a periodic
potential $U$ with minima separated by sharp maxima of height
$\phi$.The jump probability per unit time to neighbor sites
involves the Arrhenius factor: $W \exp(- \beta \,\phi)$ with
$\beta = (k_B T)^{-1}$, where $k_B$ is Boltzmann's constant
and $T$ is the temperature.
%
%
\begin{figure}
\psfig{file=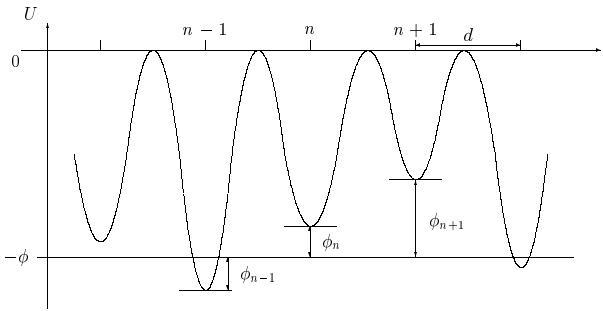,width=80mm}
\caption{
Disordered potential in the chain. $d$ is the distance between
nearest-neighbor sites. The particle jumps to neighboring sites
through potential barriers equal to $\phi - \phi_n$, where $\phi_n$
are independent identically distributed random variables with zero
mean.
}
\label{FigPot}
\end{figure}
%
In the presence of disorder, the transition probabilities are still
symmetrical but depend on the site [see Fig.~(\ref{FigPot})]:
$w_n = W \exp[- \beta (\phi-\phi_n)]$, where
$\langle \phi_n \rangle = 0$, $|\phi_n| \ll \phi$, and we suppose
that the random potential is smaller than the thermal energy, i.e.,
$\beta |\phi_n| \ll 1$. Now, suppose that an external field $E$
(toward right) is added, then the force $q \,E$ is applied on the
particle of ``charge'' $q$. The presence of $E$ alters the heights
of potential barriers. The new jump probabilities are nonsymmetric,
\begin{equation}
w_n^{\pm} = W \exp[- \beta (\phi - \phi_n \mp q \,E \,d/2)] \,,
\label{JumpT}
\end{equation}
where $d$ is the separation between neighbor sites.
In the low field limit, we can use the aproximation
$\exp[- \beta (- \phi_n \mp q \,E \,d/2)] \approx
1 + \beta \,\phi_n \pm \beta \,q \,E \,d/2$,
thus [see Eq.~(\ref{model})],
\begin{subequations}
\begin{eqnarray}
a &=& W \exp(- \beta \,\phi)
\left( 1 + \beta \,q \,E \,d/2 \right) \,,
\\
b &=& W \exp(- \beta \,\phi)
\left( 1 - \beta \,q \,E \,d/2 \right) \,,
\\
\xi_n &=& W \exp(- \beta \,\phi) \,\beta \,\phi_n \,.
\end{eqnarray}
\label{abxT}
\end{subequations}
Hence, $\gamma = b/a \approx 1 - \beta \,q \,E \,d$, i.e.,
$\epsilon = \beta \,q \,E \,d$.

Defining $a_0 = W \exp(- \beta \,\phi)$, we can write
$a = a_0 (1 + \epsilon/2)$, $b = a_0 (1 - \epsilon/2)$,
and $\xi_n = a_0 \beta \phi_n$.
Therefore, we must adapt the equations of Sec.~\ref{sec:weak}
to incorporate the lineal dependence of $a$ on $\epsilon$.
In the following, it will be useful to define the quantities
$\overline{w} = a_0 + \xi_n$,
$\overline{\beta_k} = \left< \overline{w}^{-k} \right>$, and
$\overline{\cal F} = \overline{\beta_2}/\overline{\beta_1}^2 - 1$
which are independent of $\epsilon$.
From Eqs.~(\ref{LambdaS}) and Eqs.~(\ref{LambdaR}) results
that Eqs.~(\ref{Lambda}) are valid with the substitution
$a \rightarrow a_0$. Additionally, we can see that
$\langle \psi_n \rangle_T = \overline{\beta}_1^{-1} - a_0
- a_0 \,\overline{\cal F} \,\theta_n \,\epsilon$,
where $\theta_n$ is given by Eq.~(\ref{theta}).
Now, using that
${\cal K}^b \approx a_0 \left[ {\cal K}^0
- ({\cal E}^+ - {\cal E}^-) \,\epsilon / 2 \right]$,
the explicit form of the MFPT's equation up to order $\epsilon$ is
\begin{equation}
- 1 = \overline{\beta}_1^{-1} \, {\cal K}^0 \,\langle T_n \rangle
- a_0 \left[ \overline{\cal F} \,\theta_n \, {\cal K}^0
+ \frac{1}{2} \,({\cal E}^+ - {\cal E}^-) \right]
\epsilon \,\langle T_n \rangle \,.
\label{MFPTeq2}
\end{equation}
Proposing the solution given by Eq.(\ref{SolMFPT}), we now obtain
\begin{eqnarray}
\langle T_n \rangle &\approx&
\frac{(L+1)^2 - n^2}{2 \,\overline{\beta}_1^{-1}}
\nonumber \\
&& \times
\left[ 1 + \left( -\frac{n}{3} +
\frac{L+1-n/3}{2 \,(L+1)} \,\overline{\cal F} \right)
a_0 \overline{\beta}_1 \,\epsilon \right] .
\nonumber \\
&&
\label{MFPT2}
\end{eqnarray}
In order to obtain the explicit dependence on temperature,
we additionally must replace the disorder-dependent quantities
according to
$\overline{\beta}_1 \approx
(1 + \beta^2 \langle \phi_n^2 \rangle) / a_0$
and
$\overline{\cal F} \approx \beta^2 \langle \phi_n^2 \rangle$.

On the other hand, the MRT's equation up to order $\epsilon$
results
\begin{equation}
- R_n(0) =
\overline{\beta}_1^{-1} \, {\cal K}^0 \,\langle \tau_n \rangle
+ \frac{a_0}{2} \,\epsilon \,({\cal E}^+ - {\cal E}^-)
\,\langle \tau_n \rangle \;.
\label{MRTeq2}
\end{equation}
Again, this equation is equal to that corresponding to a
nondisordered chain for small bias with transition rates
given by the substitutions
\begin{eqnarray}
a &\rightarrow& \overline{\beta}_1^{-1} + \frac{a_0}{2} \,\epsilon,
\nonumber \\
b &\rightarrow& \overline{\beta}_1^{-1} - \frac{a_0}{2} \,\epsilon,
\nonumber \\
\epsilon &\rightarrow& a_0 \,\overline{\beta}_1 \,\epsilon \,,
\label{map2}
\end{eqnarray}
Thus, up to order $\epsilon$, MRT  is equal to the expression
given by Eq.~(\ref{MRT}) with the changes $a \rightarrow a_0$
and $\beta_1 \rightarrow \overline{\beta}_1$.

\section{Conclusions}
\label{sec:Fin}

We have presented a unified framework for the FPT and RT statistics
in finite disordered chains with bias. Exact equations for the
quantities averaged over disorder were obtained for both problems
and its solutions up to first order in the bias parameter were
constructed retaining the full dependence on the system's size
and the initial condition.

We have studied the FPT and RT problems for three models of disorder.
For weak disorder, the inverse moments of the transition
probabilities are finite, and we get that the bias becomes a
control parameter for the MFPT, coupled with the fluctuation
of the disorder. The MRT is only defined in the presence of bias,
and for weak disorder the MRT's expressions are obtained from
the nondisordered case renormalizing the transition constants.
For strong disordered cases, for which the MFPT is not defined in
unbiased chains, the bias allows us to study the divergent behavior.
Amazingly, the exponents of the divergences in MFPT obtained for
vanishing bias coincide with that obtained for the averaged survival
probability in the long time regime.  The MRT is a divergent
quantity under strong disorder because the strength parameter is
not present in the corresponding equation in the small bias limit,
and the MRT is not defined for unbiased walks.

We complete the work with three appendixes.
In Appendix~\ref{app:Order} the derivation of survival and residence
probabilities for nondisordered chains is completely developed.
Apppendix~\ref{app:Green} is devoted to the detailed calculation
of Green's function in a chain with bias. Exact expressions
are given, which display the full dependence on the system's size,
initial condition, and bias strength.
The evaluation of the relevant Terwiel's cumulants is reported
in Appendix~\ref{app:Terw}.

\begin{acknowledgments}
This work was partially supported by grant from
``Se\-cre\-ta\-r\'\i a de Cien\-cia y Tec\-no\-lo\-g\'\i a de la
Uni\-ver\-si\-dad Na\-cio\-nal de C\'or\-doba 
(Code: 05/B160, Res.\ SeCyT 194/00).
\end{acknowledgments}

\appendix
\section{Survival and Residence Probabilities
in a nondisordered biased chain}
\label{app:Order}

We address the question about the survival and residence
probabilities for a RW in the finite interval $\bar{D}=[-M,L]$ of a
homogeneous chain.  In the absence of disorder, the Laplace transform 
of the evolution equation gives
$s \hat{F}_n(s) -  F_n(0) = {\cal K}^b \hat{F}_n(s)$.
This equation can be written as
\begin{equation}
\left[ a \;{\cal E}^+ + b \;{\cal E}^- - (s+a+b) \;{\cal I} \right]
\hat{F}_n(s) = - F_n(0) \;,
\label{EqF}
\end{equation}
and must be solved with the boundary conditions
corresponding to each problem,
\begin{subequations}
\begin{eqnarray}
\hat{S}_{-(M+1)}(s) = \hat{S}_{(L+1)}(s) = 0 \,,
\label{Sboundary2}
\\
\hat{R}_{n}(s) \rightarrow 0 \mbox{ for } |n| \rightarrow \infty \,,
\label{Rboundary2}
\end{eqnarray}
\label{Fboundary2}
\end{subequations}
$\forall s > 0$.
Defining the vector $A$ such that $A_{n+1} \equiv \hat{F}_n(s)$,
Eq.~(\ref{EqF}) can be written as
\begin{equation}
\left( a \;({\cal E}^+)^2 - (s+a+b) \;{\cal E}^+
+ b \;{\cal I} \right) A_n = - F_n(0) \;.
\label{EqA}
\end{equation}
Equation~(\ref{EqA}) is a second-order linear difference equation.
A particular solution for the inhomogeneous equation is the constant
$A_n = 1/s$. Proposing $A_n = x^n$ as a solution of the corresponding
homogeneous equation, it gives
$x^2 - (r+1+\gamma) \,x + \gamma = 0$, where $r=s/a$ and
$\gamma=b/a<1$. The roots of this second-order equation are given by
\begin{equation}
x_{1,2} = \frac{1}{2}
\left[
r + 1 + \gamma \pm \sqrt{(r + 1 + \gamma)^2 - 4 \,\gamma}
\right]
\label{x12}
\end{equation}
and satisfy $x_1 \,x_2 = \gamma$ with $x_1 \geq 1$.
Then, the general solution of Eq.~(\ref{EqF}) can be written as
\begin{equation}
\hat{F}_n(s) = \frac{1}{s} + c_1 \,x_1^{n} + c_2 \,x_2^{n}
\;,\;\; n \in \bar{D} \,.
\label{SolG}
\end{equation}
For the survival problem the constants $c_1$ and $c_2$ are fixed by
imposing the boundary conditions given by Eq.~(\ref{Sboundary2}).
Thus, we found for the survival probability the following expression:
\begin{widetext}
\begin{equation}
\hat{S}_n(s) = \frac{1}{s}
\left(
1 - \frac{
\left( x_1^{L+M+2} - \gamma^{L+M+2} \right) \,x_1^{n-L-1} +
\left( \gamma^{L+M+2} - x_2^{L+M+2} \right) \,x_2^{n-L-1}
}{x_1^{L+M+2} - x_2^{L+M+2}}
\right) \;.
\label{Sord}
\end{equation}
\end{widetext}
This equation is invariant under the transformation
$x_1 \leftrightarrow x_2$. From Eq.~(\ref{Sord}) and
$\hat{f}_{n}(s) = 1 - s \, \hat{S}_{n}(s)$,
we can immediately obtain the FPTD for the ordered chain.

For the residence problem, the boundary
conditions~(\ref{Rboundary2}) impose that
\begin{equation}
\hat{R}_n(s) = \left\{
\begin{array}{ll}
d_1 \,x_1^{n} & \mbox{for $n < -M$}\\
d_2 \,x_2^{n} & \mbox{for $n >  L$} \;.
\end{array}
\right.
\label{SolOut}
\end{equation}
Now, the constants $c_1$, $c_2$, $d_1$, and $d_2$ are fixed by
writing explicitly Eq.~(\ref{EqF}) for the sites $n = -(M+1)$,
$-M$, $L$, and $L+1$. Thus, we finally found the expression for the
residence probability,
%
%
\begin{eqnarray}
&& \hat{R}_n(s) = \frac{1}{s \,(x_1-x_2)}
\nonumber \\
&& \times \left\{
\begin{array}{l}
(1-x_2) \,\left( x_1^{L+M+1} - 1 \right) \,x_1^{n-L}
\,, \; n < -M \\
1 - \left[ (1-x_2) \,x_1^{n-L} + (x_1-1) \,x_2^{M+1+n} \right]
, \; n \in D \\
(x_1-1) \left( 1 - x_2^{L+M+1} \right) \,x_2^{n-L}
\, \; n >  L \;.
\end{array}
\right.
\nonumber \\
&&
\label{Rord}
\end{eqnarray}
%
%
Taking the limits of Eq.~(\ref{T_Fs}) it gives, on one hand the
MFPT for the ordered chain,
\begin{equation}
T_n = \frac{L+1-n}{a(1-\gamma)} - \frac{L+M+2}{a(1-\gamma)}
\,\frac{\gamma^n - \gamma^{L+1}}{\gamma^{-(M+1)} - \gamma^{L+1}}
\;,
\label{Tord}
\end{equation}
with $-M \leq n \leq L$, and on the other hand the MRT,
\begin{equation}
\tau_n = \frac{1}{a} \left\{
\begin{array}{ll}
\displaystyle
\frac{L+M+1}{1-\gamma}
& , \; n < -M \\
\displaystyle
\frac{L-n}{1-\gamma} + \frac{1-\gamma^{n+M+1}}{(1-\gamma)^2}
& , \; -M \leq n \leq L \\
\displaystyle
\frac{1-\gamma^{L+M+1}}{(1-\gamma)^2} \;\gamma^{n-L}
& , \; n >  L \;.
\end{array}
\right.
\label{Tauord}
\end{equation}
Equations~(\ref{TordL}) and~(\ref{TauordL}) follow from
Eqs.~(\ref{Tord}) and and~(\ref{Tauord}), respectively,
taking $M=L$.

\section{Green's Function in a chain with Bias}
\label{app:Green}

In this section we are concerned with the solution of
Eq.~(\ref{EqG}). In components, this backward equation
can be written as
\begin{equation}
a \,\hat{G}_{n+1 m}(s) + b \,\hat{G}_{n-1 m}(s)
- (s+a+b) \,\hat{G}_{n m}(s) = -\delta_{n m} .
\label{EqG2}
\end{equation}
We must solve this equation with the boundary conditions
corresponding to each problem,
\begin{subequations}
\begin{equation}
\hat{G}^S_{-(M+1) \,m}(s) = \hat{G}^S_{(L+1) \,m}(s) = 0
\;\forall \,s \mbox{ and }  m \in \bar{D} \;,
\label{GSboundary2}
\end{equation}
\begin{equation}
\hat{G}^R_{n m}(s) \rightarrow 0 \;
\mbox{ for } |n| \rightarrow \infty
\mbox{ and } m \mbox{ finite} \;.
\label{GRboundary2}
\end{equation}
\label{Gboundary2}
\end{subequations}
Here, the superscript~$S$ ($R$) denotes the solution corresponding 
to the survival (residence) problem.

The solution of Eq.~(\ref{EqG2}) in a finite domain (survival
problem) can be constructed using the method of images~\cite{MW87}.
It consists in summing to the Green's function in the absence of
boundaries, terms corresponding to the specular image of the
index $n$ with respect to the boundary considered. For absorbing
boundaries the image must have a negative sign. Moreover,
in the presence of bias, the image must change the bias direction
($a \leftrightarrow b$).  Additionally, in a closed domain, each
boundary reflects the image of the other boundary.
This fact introduces an infinite set of images.
However, we adopt here the simpler algebraic approach developed
in Appendix~\ref{app:Order}. Thus, we propose
\begin{equation}
\hat{G}^S_{n m}(s) = \left\{
\begin{array}{ll}
\gamma_1 \,x_1^{n} + \gamma_2 \,x_2^{n} & \mbox{for $n \leq m$}\\
\delta_1 \,x_1^{n} + \delta_2 \,x_2^{n} & \mbox{for $n \geq m$}
\end{array}
\right.
\mbox{ with } n, m \,\in \bar{D} \;,
\label{SolGS}
\end{equation}
where the functions $x_{1,2}$ are defined in Eq.~(\ref{x12}).
The constants $\gamma_1$, $\gamma_2$, $\delta_1$, and $\delta_2$
are solutions of the set of four algebraic equations that result
from imposing the boundary conditions given by
Eq.~(\ref{GSboundary2}), imposing the equality between both
expressions given by Eq.~(\ref{SolGS}) for $n=m$, and writing
Eq.~(\ref{EqG2}) for $n=m$. Additionally, using the relation
$a \,x_{1,2}^2 - (s+a+b) \,x_{1,2} + b = 0$ we obtain
\begin{widetext}
\begin{equation}
\hat{G}^S_{n m}(s) = \frac{1}{D} \,\left\{
\begin{array}{ll}
x_1^{n+M+1} \left( x_1^{-m+L+1} - x_2^{-m+L+1} \right)
+
x_2^{n+M+1} \left( x_2^{-m+L+1} - x_1^{-m+L+1} \right)  
& , n \leq m
\\
x_1^n \,x_2^{L+1} 
\left( x_1^{-m} x_2^{M+1} - x_2^{-m} x_1^{M+1} \right)
+
x_2^n \,x_1^{L+1} 
\left( x_2^{-m} x_1^{M+1} - x_1^{-m} x_2^{M+1} \right)  
& , n \geq m \;,
\end{array}
\right.
\label{GS}
\end{equation}
\end{widetext}
where
$D = a \,(x_1 - x_2) \,\left( x_1^{L+M+2} - x_2^{L+M+2} \right)$.
Defining the variable $z = s+a+b - 2 \,\mu$,
the constant $\mu = \sqrt{ab}$, and the functions
\begin{subequations}
\begin{equation}
A(z) = 1 + \frac{z}{2 \mu} -
\left( \frac{z}{\mu} + \frac{z^2}{4 \mu^2}  \right)^{1/2} \;,
\end{equation}
\begin{equation}
B(z) = \frac{1}{2 \mu} \,
\left( \frac{z}{\mu} + \frac{z^2}{4 \mu^2}  \right)^{-1/2} \;,
\end{equation}
\label{AB}
\end{subequations}
results in 
$x_2 = \sqrt{\gamma} \,A(z)$ and $a (x_1 - x_2) = B(z)^{-1}$.
Using these functions we can recast the
expression~(\ref{GS}), for the case $M=L$, as
\begin{eqnarray}
\hat{G}^S_{n m}(s) &=& \gamma^{\frac{n-m}{2}} \,B(z)
\,\left[ 1- A(z)^{4 \,(L+1)} \right]^{-1}
\nonumber\\
&& \times \left[ \rule{0mm}{4mm} \right. A(z)^{|n-m|}
\nonumber\\
&& - A(z)^{2 \,(L+1)}
\left( A(z)^{n+m} + A(z)^{-(n+m)} \right)
\nonumber\\
&& \left. + A(z)^{4 \,(L+1)} \,A(z)^{-|n-m|} \right] \;.
\label{GS(s)}
\end{eqnarray}
The last expression satisfies the symmetry relation
$(a,b),(n,m) \leftrightarrow (b,a),(m,n)$.
Denoting by $\hat{\cal G}^S_{n m}(s)$ the Green's function
corresponding to the survival problem without bias (and taking
$\mu = \sqrt{ab}$)~\cite{HC90}, we can immediately see that
$\hat{G}^S_{n m}(s) =
\gamma^{\frac{n-m}{2}} \,\hat{\cal G}^S_{n m}(z)$.
Moreover, given that
$\left( z - \mu \,{\cal K}^0 \right) \hat{\cal G}^S(z) = \openone$,
we immediately obtain Eq.~(\ref{EqG2}).

We now compute the propagator
$J_{n m}(s) = {\cal K}^0 \hat{G}_{n m}(s)$.
Applying the operator ${\cal K}^0$ defined in
Eq.~(\ref{operators}) to Eq.~(\ref{GS(s)}),
the following expression is found:
\begin{widetext}
\begin{eqnarray}
J^S_{n m}(s) &=& B(z) \,\left[ 1- A(z)^{4 \,(L+1)} \right]^{-1}
\left[ \rule{0mm}{5mm} \right. - A(z)^{2 \,(L+1)}
\nonumber\\
&& \times
\left\{
\left( \frac{\sqrt{\gamma}}{A(z)} \right)^n  \,
\left[ \sqrt{\gamma} \,A(z) \right]^{-m}
\left(
\frac{\sqrt{\gamma}}{A(z)} + \frac{A(z)}{\sqrt{\gamma}} - 2
\right)
\right.
\nonumber\\
&& +
\left.
\left[ \sqrt{\gamma} \,A(z) \right]^n        \,
\left( \frac{A(z)}{\sqrt{\gamma}} \right)^m
\left( \sqrt{\gamma} \,A(z)+\frac{1}{\sqrt{\gamma} \,A(z)}-2 \right)
\right\}
+ {\cal Z}_{n m} \;
\left. \rule{0mm}{5mm} \right] \;,
\nonumber\\
&&
\label{JS}
\end{eqnarray}
where $n, m \,\in D$ and
\begin{equation}
{\cal Z}_{n m} =
\left\{
\begin{array}{ll}
\displaystyle
\left( \frac{\sqrt{\gamma}}{A(z)} \right)^{n-m-1} \,
\left( \frac{\sqrt{\gamma}}{A(z)} - 1 \right)^2
+
A(z)^{4 \,(L+1)} \,
\left( \sqrt{\gamma} \,A(z) \right)^{n-m-1} \,
\left( \sqrt{\gamma} \,A(z) - 1 \right)^2
&, n < m \\
\displaystyle
\sqrt{\gamma} \,A(z) + \frac{A(z)}{\sqrt{\gamma}} - 2
+
A(z)^{4 \,(L+1)} \,
\left(
\frac{\sqrt{\gamma}}{A(z)} + \frac{1}{\sqrt{\gamma} \,A(z)} - 2
\right)
&, n = m \\
\displaystyle
\left( \sqrt{\gamma} \,A(z) \right)^{n-m-1} \,
\left( \sqrt{\gamma} \,A(z) - 1 \right)^2
+
A(z)^{4 \,(L+1)} \,
\left( \frac{\sqrt{\gamma}}{A(z)} \right)^{n-m-1} \,
\left( \frac{\sqrt{\gamma}}{A(z)} - 1 \right)^2
&, n > m \;.
\end{array}
\right.
\label{Z}
\end{equation}
\end{widetext}

On the other hand, for the residence problem we propose
\begin{equation}
\hat{G}^R_{n m}(s) = \left\{
\begin{array}{ll}
\lambda_1 \,x_1^{n} & \mbox{for $n \leq m$}\\
\lambda_2 \,x_2^{n} & \mbox{for $n \geq m$}
\end{array}
\right.
\mbox{ with } n, m \,\in \bar{D} \;.
\label{SolGR}
\end{equation}
This expression immediately satisfies the homogeneous case of
Eq.~(\ref{EqG2}) and the boundary conditions~(\ref{GRboundary2}),
given that $x_1 \,x_2 = \gamma < 1$ with $x_1 \geq 1$.
The constants $\lambda_1$ and $\lambda_2$ are fixed by imposing
for $n=m$, the equality between both expressions in Eq.~(\ref{SolGR}),
and writting Eq.~(\ref{EqG2}) for $n=m$.
Hence, we obtain
\begin{equation}
\hat{G}^R_{n m}(s) = \frac{1}{a \,(x_1-x_2)} \,\left\{
\begin{array}{ll}
\gamma^{n-m} \,x_2^{m-n}
& \mbox{for } n \leq m \\
x_2^{n-m}
& \mbox{for } n \geq m \;.
\end{array}
\right.
\label{GR}
\end{equation}
Using the functions defined by Eq.~(\ref{AB}), the last expression
can be recast as
\begin{equation}
\hat{G}^R_{n m}(s) = \gamma^{(n-m)/2} \,B(z) \,A(z)^{|n-m|} \;.
\label{GR(z)}
\end{equation}
Applying the operator ${\cal K}^0$ to Eq.~(\ref{GR(z)})
we find:
\begin{eqnarray}
J^R_{n m}(s) &=& \gamma^{(n-m)/2} \,B(z) \;A(z)^{|n-m|}
\nonumber\\
&& \times
\left\{
\begin{array}{ll}
\gamma^{1/2} \,A(z)^{-1} + \gamma^{-1/2} \,A(z)      - 2 &, n < m \\
\gamma^{1/2} \,A(z)      + \gamma^{-1/2} \,A(z)      - 2 &, n = m \\
\gamma^{1/2} \,A(z)      + \gamma^{-1/2} \,A(z)^{-1} - 2 &, n > m .
\end{array}
\right.
\nonumber\\
\label{JR}
\end{eqnarray}

We are interested in analyzing the propagator $J_{n m}(s)$ in the
limit $s \rightarrow 0$ for a given bias, also we are concerned with
the expansion for small bias ($\epsilon \rightarrow 0$). We must
stress that these limits do not commute. Thus, we first must take the
limit $s \rightarrow 0$ for a fixed bias, and then perform the
expansion in the parameter $\epsilon$.
Defining the variable $y = s/(a-b)$, we get from Eq.~(\ref{AB}),
for small $s$, that
\begin{subequations}
\begin{equation}
A(z) \approx \sqrt{\gamma}
\left( 1 - y + \frac{1}{1-\gamma} \,y^2  \right)
\,,
\end{equation}
\begin{equation}
B(z) \approx \frac{1}{a \,(1-\gamma)}
\left( 1 - \frac{1+\gamma}{1-\gamma} \,y +
\frac{(1+\gamma)^2+2 \gamma}{(1-\gamma)^2} \,y^2  \right)
\,.
\end{equation}
\label{AB(y)}
\end{subequations}
These approximations allow us to write the propagator in the form:
$J_{n m}(s) \approx \Lambda_{n m} + {\cal O}(y)$,
where for the survival problem
\begin{eqnarray}
&& \Lambda^S_{n m} = \frac{1}{a \,(1-\gamma)} \,
\left( 1 - \gamma^{2 \,(L+1)}  \right)^{-1}
\nonumber \\
&& \times
\left\{
\begin{array}{ll}
(1-\gamma^2) \,\left( -\gamma^{n+L} +\gamma^{n-m+2L+1} \right)
&, n < m \\
-(1-\gamma^2) \,\gamma^{n+L}+(1-\gamma) \left( \gamma^{2L+1}-1 \right)
&, n = m \\
(1-\gamma^2) \,\left( -\gamma^{n+L} +\gamma^{n-m-1} \right)
&, n > m ,
\end{array}
\right.
\nonumber \\
&&
\label{LambdaS}
\end{eqnarray}
and, on the other hand, for the residence problem
\begin{equation}
\Lambda^R_{n m} = \frac{1}{a}
\left\{
\begin{array}{ll}
0
&, n < m \\
-1
&, n = m \\
(1-\gamma) \, \gamma^{n-m-1}
&, n > m \;.
\end{array}
\right.
\label{LambdaR}
\end{equation}
We remark that we keep the exact dependence in the parameter $L$
(size of the system) for the FPT problem. From Eq.~(\ref{LambdaS}),
we can also see that the $s$-independent contribution is quite
different from that which is obtained without bias~\cite{HC90}.
In the presence of bias we get nondiagonal contributions to the
propagator for the survival and residence problems in the limit
$s \rightarrow 0$, which also remain for small bias, 
as has been shown in Eq.~(\ref{Lambda}).

\section{Terwiel's Cumulants}
\label{app:Terw}

The Terwiel's cumulants were introduced in Ref.~\cite{Terwiel}.
In this Appendix we are concerned with the evaluation of the
cumulants of the random operator $\psi_n$ defined by
\begin{equation}
\psi_n = M_n \left( 1 - N_n \,{\cal P} M_n \right) \,,
\label{psi3}
\end{equation}
where 
\begin{equation}
M_n = \frac{\xi_n}{1 - \xi_n \,J_{n n}}
\label{M2}
\end{equation}
is a random variable, and
\begin{equation}
N_n = \frac{J_{n n}}{1 - \langle M_n \rangle \,J_{n n}} \,.
\label{N}
\end{equation}
From its definition,
\begin{equation}
\langle \psi_n \psi_{m_1} \ldots \psi_{m_p} \rangle_T =
{\cal P} \,\psi_{n} \,(1 - {\cal P}) \,\psi_{m_1}
\dots (1 - {\cal P}) \,\psi_{m_p}
\label{T}
\end{equation}
can be easily obtained
\begin{eqnarray}
\langle \psi_n \rangle_T &=& \langle \psi_n \rangle
\,, \nonumber \\
\langle \psi_n \psi_m \rangle_T &=&
\langle \psi_n \psi_m \rangle -
\langle \psi_n \rangle \langle \psi_m \rangle
\,, \nonumber \\
\langle \psi_n \psi_{m_1} \psi_{m_2} \rangle_T &=&
\langle \psi_n \psi_{m_1} \psi_{m_2} \rangle -
\langle \psi_n \rangle \langle \psi_{m_1} \psi_{m_2} \rangle
\nonumber \\
&&-
\langle \psi_n \psi_{m_1} \rangle \langle \psi_{m_2} \rangle +
\langle \psi_n \rangle \langle \psi_{m_1}
\rangle \langle \psi_{m_2} \rangle
\,. \nonumber \\
\label{T123}
\end{eqnarray}
For $n \neq m$, $\psi_n$ and $\psi_m$ are statistically independent,
then $\langle \psi_n \psi_m \rangle =
\langle \psi_n \rangle \langle \psi_m \rangle$.
Therefore,
\begin{eqnarray}
\langle \psi_n \psi_m \rangle_T &=& 0
\,, \nonumber \\
\langle \psi_n \psi_m \psi_{n'} \rangle_T &=& 0 \;\;(n \neq n')
\,, \nonumber \\
\langle \psi_n \psi_m \psi_n \rangle_T &=&
\langle \psi_n \psi_m \psi_n \rangle +
\langle \psi_n \rangle \langle \psi_m
\rangle \langle \psi_n \rangle
\,,
\label{T23}
\end{eqnarray}
and using Eqs.~(\ref{psi3})-(\ref{N}) results
\begin{eqnarray}
\langle \psi_n  \rangle &=&
\frac{\langle M_n \rangle}
{1 + \langle M_n \rangle \, J_{n n}}
\,, \nonumber \\
\langle \psi_n \psi_m \psi_n \rangle_T &=&
\left( 1 + \langle M_n \rangle \, J_{n n} \right)^{-2}
\nonumber \\
&& \times
\left( \langle M_n^2 \rangle \langle M_m \rangle
- N_m \langle M_n \rangle^2 \langle M_m \rangle^2 \right)
\,. \nonumber \\
\label{T13}
\end{eqnarray}

In the limit $s \rightarrow 0$, from Eq.~(\ref{Lambda}) the
diagonal components of the propagator $J_{n n}$ can be written as
\begin{equation}
J_{n n} \approx - \frac{1}{a} \left( 1 + \theta_n \epsilon \right)
\,,
\label{J}
\end{equation}
where
\begin{equation}
\theta_n =
\left\{
\begin{array}{ll}
\displaystyle\frac{L+1-n}{2 \,(L+1)} & \,, \mbox{FPT problem} \\
0 & \,, \mbox{RT problem} \;.
\end{array}
\right.
\label{theta}
\end{equation}
Using the transfer rate $\omega = a + \xi_n$, we obtain
\begin{equation}
\left( 1 - \xi_n J_{n n} \right)^{-1} \approx
\frac{a}{\omega} \left[ 1 - \left( 1 - \frac{a}{\omega} \right)
\theta_n \, \epsilon \right] \,.
\label{-1}
\end{equation}
Thus, we can write
\begin{equation}
M_n \approx a \left( 1 - \frac{a}{\omega} \right)
\left[ 1 - \left( 1 - \frac{a}{\omega} \right)
\theta_n \, \epsilon \right] \,.
\label{Mn}
\end{equation}

For a disorder of class~(\ref{weak}) for which the quantities
$\beta_k$ are finite, results in
\begin{eqnarray}
\left< \left( 1 - \frac{a}{\omega} \right) \right> &=&
1 - a \,\beta_1
\,, \nonumber \\
\left< \left( 1 - \frac{a}{\omega} \right)^2 \right> &=&
1 - 2 a \,\beta_1 + a^2 \,\beta_2
\,.
\label{aux}
\end{eqnarray}
Hence,
\begin{eqnarray}
\langle M_n \rangle &\approx&
a \left[ 1 - a \,\beta_1 -
\left( 1 - 2 a \,\beta_1 + a^2 \,\beta_2 \right)
\theta_n \,\epsilon \right]
\,, \nonumber \\
1 + \langle M_n \rangle J_{n n} &\approx& a \,\beta_1
\left[ 1 - \left( 1 - a \,\frac{\beta_2}{\beta_1} \right)
\theta_n \,\epsilon\right]
\,, \nonumber \\
N_n &\approx& - \frac{1}{a^2} \,\beta_1^{-1}
\left[ 1 + \left( 2 - a \,\frac{\beta_2}{\beta_1} \right)
\theta_n \,\epsilon\right]
\label{approx}
\end{eqnarray}
and therefore
\begin{eqnarray}
\langle \psi_n \rangle &=&
\beta_1^{-1} - a - a \,{\cal F} \,\theta_n \,\epsilon
\,, \nonumber \\
\langle \psi_n \psi_m \psi_n \rangle_T &=&
\beta_1^{-2}
\Big[
a \left( 1 - a \,\beta_1 \right)
\left( 1 - 2 a \,\beta_1 + a^2 \,\beta_2 \right)
\nonumber \\
&&+
\beta_1^{-1} \left( 1 - a \,\beta_1 \right)^4
\Big]
+ {\cal O}(\epsilon)
\,,
\label{T13weak}
\end{eqnarray}
for $n \neq m$. Here, we have used that
${\cal F} = \left( \beta_2 / \beta_1^2 \right) - 1$.

For the strong classes of disorder, the quantities $\beta_k$ diverge
in the limit $\epsilon \rightarrow 0$ and can be seen that
$\left< \left( 1 - \displaystyle\frac{a}{\omega} \right)^k \right>
\propto \beta_k$.
In particular, for disorder of class~(\ref{windows})
$\beta_1 \propto |\ln \epsilon|$ and
$\beta_k \propto \epsilon^{1-k}$ for $(k>1)$.
Thus, from Eq.~(\ref{Mn}) we obtain
$\langle M_n \rangle \propto |\ln \epsilon|$ and
$\langle M_n^2 \rangle \propto \epsilon^{-1}$,
which give
\begin{eqnarray}
\langle \psi_n  \rangle \approx \beta_1^{-1} &\propto&
\frac{1}{|\ln \epsilon|}
\,, \nonumber \\
\langle \psi_n \psi_m \psi_n \rangle_T &\propto&
\frac{1}{\epsilon \,|\ln \epsilon|}
\,.
\label{T13win}
\end{eqnarray}
Finally, for a disorder of class~(\ref{alpha}),
$\beta_k \propto \epsilon^{1-k-\alpha}$ and results in
$\langle M_n \rangle \propto \epsilon^{-\alpha}$ and
$\langle M_n^2 \rangle \propto \epsilon^{-1-\alpha}$.
Therefore,
\begin{eqnarray}
\langle \psi_n  \rangle \propto \beta_1^{-1} &\propto&
\epsilon^{\alpha}
\,, \nonumber \\
\langle \psi_n \psi_m \psi_n \rangle_T &\propto&
\epsilon^{-1}
\,.
\label{T13mc}
\end{eqnarray}
%

%


\begin{thebibliography}{10}
\bibitem{note:onestep} We are considering only one-step processes.
For general processes with large jumps allowed the boundary condition
must be extended over all sites $n_0$ out of $D$.
\bibitem{note:MFPT} From Eq.~(\ref{T_f}) the general expression
for MFPT is given by
$T_{n_0} = \lim_{s \rightarrow 0}
\left[ \hat{S}_{n_0}(s) + s \,\hat{S}'_{n_0}(s) \right]$.
\bibitem{note:multipl} A related interesting model is the
multiplicative one: $w_n^+ = a \,\xi_n$ and $w_n^- = b \,\xi_n$,
where $\langle \xi_n \rangle = 1$.
Thus, $w_n^- / w_n^+ = 1- \epsilon$ and the corresponding
evolution equation that results from Eq.~(\ref{EqFt}) is
$s \hat{F}_n(s) - F(0) = \xi_n \, {\cal K}^b \,\hat{F}_n(s)$.
Here, we can also obtain the symmetrical disordered system and
the ordered biased chain as limit cases.
\bibitem{note:MRT} We must note that from Eqs.~(\ref{resid})
and (\ref{tau_R}), the MRT is defined only if $P(n,t|n_0,t_0)$
is a integralable function in the interval $(0, \infty)$.
This condition is only fulfilled in the presence of bias.
\end{thebibliography}
\end{document}